\documentclass[prd,twocolumn,showpacs,preprintnumbers,amsmath,nofootinbib,amssymb,floatfix]{revtex4-1}
\usepackage{graphicx}
\usepackage{epsfig,float}
\usepackage[sort&compress]{natbib}
\usepackage{subfigure}
\usepackage{amsmath}
\usepackage{amsfonts}
\usepackage{cancel}
\usepackage{lmodern,dsfont}
\usepackage{amssymb}
\usepackage{hyperref}
\begin{document}
\arraycolsep1.5pt
\newcommand{\Ima}{\textrm{Im}}
\newcommand{\Rea}{\textrm{Re}}
\newcommand{\mev}{\textrm{ MeV}}
\newcommand{\gev}{\textrm{ GeV}}
\newcommand{\dtres}{d^{\hspace{0.1mm} 3}\hspace{-0.5mm}}
\newcommand{\rts}{ \sqrt s}
\newcommand{\non}{\nonumber \\[2mm]}
\newcommand{\eps}{\epsilon}
\newcommand{\half}{\frac{1}{2}}
\newcommand{\thalf}{\textstyle \frac{1}{2}}
\newcommand{\Nmass}{M_{N}} 
\newcommand{\delmass}{M_{\Delta}} 
\newcommand{\pimass}{\mu}  
\newcommand{\rhomass}{m_\rho} 
\newcommand{\piNN}{f}      
\newcommand{\rhocoup}{g_\rho} 
\newcommand{\fpi}{f_\pi} 
\newcommand{\f}{f} 
\newcommand{\nucfld}{\psi_N} 
\newcommand{\delfld}{\psi_\Delta} 
\newcommand{\fpiNN}{f_{\pi N N}} 
\newcommand{\fpiND}{f_{\pi N \Delta}} 
\newcommand{\GMquark}{G^M_{(q)}} 
\newcommand{\vecpi}{\vec \pi}
\newcommand{\vectau}{\vec \tau}
\newcommand{\vecrho}{\vec \rho}
\newcommand{\delmu}{\partial_\mu}
\newcommand{\delMu}{\partial^\mu}
\newcommand{\nn}{\nonumber}
\newcommand{\bi}{\bibitem}
\newcommand{\vs}{\vspace{-0.20cm}}
\newcommand{\be}{\begin{equation}}
\newcommand{\ee}{\end{equation}}
\newcommand{\ba}{\begin{eqnarray}}
\newcommand{\ea}{\end{eqnarray}}
\newcommand{\ropi}{$\rho \rightarrow \pi^{0} \pi^{0}
\gamma$ }
\newcommand{\roeta}{$\rho \rightarrow \pi^{0} \eta
\gamma$ }
\newcommand{\omepi}{$\omega \rightarrow \pi^{0} \pi^{0}
\gamma$ }
\newcommand{\omeeta}{$\omega \rightarrow \pi^{0} \eta
\gamma$ }
\newcommand{\ul}{\underline}
\newcommand{\del}{\partial}
\newcommand{\rth}{\frac{1}{\sqrt{3}}}
\newcommand{\rsix}{\frac{1}{\sqrt{6}}}
\newcommand{\sq}{\sqrt}
\newcommand{\fr}{\frac}
\newcommand{\pr}{^\prime}
\newcommand{\ov}{\overline}
\newcommand{\Gm}{\Gamma}
\newcommand{\rw}{\rightarrow}
\newcommand{\rgl}{\rangle}
\newcommand{\De}{\Delta}
\newcommand{\Dp}{\Delta^+}
\newcommand{\Dm}{\Delta^-}
\newcommand{\Dz}{\Delta^0}
\newcommand{\Dpp}{\Delta^{++}}
\newcommand{\Sg}{\Sigma^*}
\newcommand{\Sp}{\Sigma^{*+}}
\newcommand{\Sm}{\Sigma^{*-}}
\newcommand{\Sz}{\Sigma^{*0}}
\newcommand{\X}{\Xi^*}
\newcommand{\Xm}{\Xi^{*-}}
\newcommand{\Xz}{\Xi^{*0}}
\newcommand{\Om}{\Omega}
\newcommand{\Omm}{\Omega^-}
\newcommand{\kp}{K^+}
\newcommand{\kz}{K^0}
\newcommand{\pip}{\pi^+}
\newcommand{\pim}{\pi^-}
\newcommand{\piz}{\pi^0}
\newcommand{\et}{\eta}
\newcommand{\kb}{\ov K}
\newcommand{\km}{K^-}
\newcommand{\kbz}{\ov K^0}
\newcommand{\ksb}{\ov {K^*}}

\def\tstrut{\vrule height2.5ex depth0pt width0pt} 
\def\jtstrut{\vrule height5ex depth0pt width0pt} 

\title{$\Lambda(1405)$ poles obtained from $\pi^0 \Sigma^0$ photoproduction data}

\author{L. Roca$^1$  and E. Oset$^2$}
\affiliation{
$^1$ Departamento de F\'\i sica, Universidad de Murcia, E-30100 Murcia, Spain.\\
$^2$ Departamento de F\'{\i}sica Te\'orica and IFIC, Centro Mixto Universidad de Valencia-CSIC,
Institutos de Investigaci\'on de Paterna, Aptdo. 22085, 46071 Valencia,
Spain
 }

\date{\today}

\begin{abstract}

We present a strategy to extract the position of the two 
$\Lambda(1405)$ poles from experimental photoproduction data
measured recently at different energies in the $\gamma p \to K^+ \pi^0
\Sigma^0 $ reaction at Jefferson Lab. By means of a chiral dynamics
motivated potential but with free parameters, we solve the Bethe
Salpeter equation in the coupled channels $\bar K N$ and $\pi \Sigma$ in
isospin I=0 and parameterize the amplitude for the photonuclear reaction
in terms of a linear combination of the  $\pi \Sigma \to \pi \Sigma$ and
$\bar K N \to \pi \Sigma$ scattering amplitudes in I=0, with a different
linear combination for each energy. Good fits to the data are obtained
with some sets of parameters, by means of which one can also predict the
cross section for the $K^- p \to \pi^0 \Sigma^0 $ reaction. These later
results help us decide among the possible solutions. The result is that
the different solutions lead to two poles  similar to those found in
the chiral unitary approach. With
the best result we find the two $\Lambda(1405)$ poles at $1385-68i\mev$
and $1419-22i\mev$.

\end{abstract}


\maketitle

\section{Introduction}
\label{Intro} 

The issue of the nature of the $\Lambda(1405)$ has captured great
 attention through the years. Very early it was already postulated that
 it could be a resonance made from the interaction of the coupled
 channels $\bar K N$ and $\pi \Sigma$
 \cite{Dalitz:1960du,Dalitz:1967fp}. Other works followed, looking at
 the $\bar{K}N$ interaction from other perspective \cite{Veit:1984an}. A
 big step forward was made possible with the use of chiral dynamics in
 its unitarized form, the so called chiral unitary approach, which has
 brought a new perspective to the problem and has shown the importance
 of coupled channels and unitarity
 \cite{Kaiser:1995eg,Kaiser:1996js,Oset:1998it,Oller:2000fj,Lutz:2001yb,Oset:2001cn,Hyodo:2002pk,cola,GarciaRecio:2002td,GarciaRecio:2005hy,Borasoy:2005ie,Oller:2006jw,Borasoy:2006sr,hyodonew}.
 One of the novel aspects of these works has been the finding of two
 poles, and  thus two states, rather than one, associated to the
 experimental peaks of the $\Lambda(1405)$ resonance. Hints of the two
 poles had been found in \cite{Fink:1989uk}, using the chiral quark
 model, and in \cite{Oller:2000fj} within the chiral unitary approach. A
 systematic search and discussion on their origin was done in
 \cite{cola}, where it was found that the two poles stem from a SU(3)
 singlet and octet of the chiral dynamical theory. One pole is around
 1390 MeV, is wide ($\Gamma>100$ MeV) and couples mostly to $\pi
 \Sigma$. The other pole appears in all these theories around 1420 MeV,
 is narrow ($\Gamma=30$ MeV) and couples mostly $\bar{K}N$.  Due to the
 existence of these two poles, the peak observed in experiments should be
 different in different reactions, as has been the case in the reactions
 studied so far
 \cite{Thomas:1973uh,Hemingway:1984pz,Niiyama:2008rt,prakhov,Moriya:2012zz,Moriya:2013eb,Zychor:2007gf,fabbietti}.
 The early experiments gave a peak around 1405 MeV, which served to give
 the nominal mass to the resonance. In view of this, new reactions were
 devised that would show the peak around 1420\mev, close to the second
 pole found in the chiral unitary approach works. The obvious thing was
 to look for reactions where the $\Lambda(1405)$ production would be
 induced by the $\bar K N$ interaction. This seems to be contradictory
 since the threshold for $\bar K N$, at 1432\mev, is higher than the
 mass of the two poles. Something should be done to remove energy from
 the initial state while still guaranteeing that the resonance was
 initiated by the $\bar K N$ channel. A first suggestion was made in 
 \cite{Nacher:1999ni}, where the radiative production of the
 $\Lambda(1405)$ resonance in $K^-$ collisions on protons was proposed.
 The photon was radiated from the incoming $K^-$ and then one still had
 the $K^- p$ state to form the resonance. Although the existence of two
 poles was not well known at that time, the theoretical cross section 
 indeed provided a
 narrow peak around 1420 MeV. Awaiting for this reaction to be done, an
 equivalent  reaction, the $K^- p \to \pi^0\pi^0 \Sigma^0$
 \cite{prakhov}, was measured and a peak was indeed seen around 1420 MeV
 and narrower than the one observed in
 \cite{Thomas:1973uh,Hemingway:1984pz}. In this case it was the $\pi^0$
 which was emitted from the initial nucleon, and one still had the $K^-
 p$ initial state, although with smaller energy, to form the resonance.
 A theoretical description of this reaction in terms of the chiral
 unitary approach of \cite{Oset:1998it} was provided in \cite{magaslam}.
 Further support for the two pole picture came from \cite{Braun:1977wd}
 where the $K^- d \to n \pi \Sigma$ reaction was measured and a neat
 peak was observed around 1420 MeV. The reaction was studied
 theoretically in \cite{sekihara} and it was found that the mechanism of
 scattering of the kaon with a neutron, losing some energy, followed by
 rescattering of the kaon with the proton to produce the
 $\Lambda(1405)$, provided the right strength and shape observed in
 experiment. It was found that kaons in flight were preferable since
 they allowed to clearly separate the peaks due to single and double
 scattering\footnote{An experiment along these lines is being proposed
 for JPARC \cite{nuomi} }. More difficult but still possible, measuring
 neutrons in coincidence, was to see the resonance signal with the kaons
 of the  DAFNE facility  \cite{Jido:2010rx}.\footnote{A recent  paper
 \cite{haidenbauer} questioned the approach of \cite{sekihara}, showing
 problems with threshold behaviour in \cite{sekihara}. In a reply to
 that work \cite{reply}, it was shown that the comments were appropriate
 under the choice of the kinetic energy for $H_0$ in the deuteron
 Hamiltonian $H=H_0+V$, but this is not necessary nor convenient in the
 multiple scattering expansion, and the Watson expansion provided an
 alternative where the interaction of the nucleons in the deuteron was
 taken into account, leading to the approach of \cite{sekihara}.}

  As we can see, there is mounting evidence of a state around 1420 MeV,
complementing another state more around 1400 MeV. Most reactions get
contribution from both poles, but some, as those discussed above, give
more weight to the pole around 1420 MeV, producing then a peak around
this energy. 

  The surprising thing is that the theoretical approaches dealing with the $\bar K N$ interaction and predicting the properties of the $\Lambda(1405)$ have paid 
little or null attention to the reactions where the resonance is produced. One of the exceptions to this rule is the model constructed for photoproduction of the $\Lambda(1405)$, done in \cite{Nacher:1998mi} before the experiment was performed, which predicted the basic features and strength of the reaction. Similarly, the 
$\pi^- p \to K^0 \pi\Sigma$ reaction of \cite{Thomas:1973uh} was studied theoretically in \cite{Hyodo:2003jw}, the $p p \to p K^+ \pi \Sigma$ reaction of \cite{Zychor:2007gf} in \cite{Geng:2007vm}, the $K^- p \to \pi^0\pi^0 \Sigma^0$ of \cite{prakhov} in \cite{magaslam} and the $K^- d \to n \pi \Sigma$ reaction of \cite{Braun:1977wd} in \cite{sekihara}. The chiral unitary approach with the potential from the lowest order chiral Lagrangians was used in all these studies. Meanwhile more refined models have been developed \cite{Borasoy:2005ie,Oller:2006jw,Borasoy:2006sr,hyodonew} that contain the next to leading order terms in the potential. It is, however, interesting to observe that the results of \cite{Oset:1998it} with the lowest order potential provide all the observables on cross sections and threshold ratios within the error bands provided by the more refined theoretical potential of \cite{Borasoy:2006sr}.

   Other theoretical works do not conduct a thorough search of these
reactions but try to be consistent with the data of $\Lambda(1405)$
production commenting that with a reaction amplitude made out from
linear combinations of the $\bar K N \to \pi \Sigma $ and $\bar K N \to
\bar K N$ amplitudes one could in principle obtain consistent shapes for
the $\pi \Sigma$ mass distributions where the $\Lambda(1405)$ is always
found. This is the case of \cite{Oller:2000fj,Oller:2006jw,hyodonew}. In
\cite{Oller:2000fj} one goes even further since such a test is demanded in
the fit to the data and in \cite{Oller:2006jw} even the $K^- p \to
\pi^0\pi^0 \Sigma^0$ reaction of \cite{prakhov} is demanded to be
reproduced using the theoretical model of \cite{magaslam}. This comment
is most appropriate, since in recent times, apart from the valuable data
for the $K^- p \to K^- p$ amplitude at threshold of the SIDHARTA
experiment \cite{Bazzi:2011zj}, no more data on $\bar K N$ induced
reactions have been produced. This contrasts with the mounting
experimental data on reactions producing the $\Lambda(1405)$
\cite{Niiyama:2008rt,prakhov,Moriya:2012zz,Moriya:2013eb,Zychor:2007gf,fabbietti}.

     In the present paper we would like to give a step in the direction
of showing the value of the $\Lambda(1405)$ production reactions to get
an insight on the properties of the $\Lambda(1405)$ states and $\bar K
N$ scattering. For this purpose we have taken all the data on
photoproduction of $\Lambda(1405)$ at different energies of the CLAS
collaboration at Jefferson Lab \cite{Moriya:2012zz,Moriya:2013eb}, with $\pi^0
\Sigma^0$ in the final state, and have performed a fit to these data in
terms of linear combinations of the  $\bar K N \to \pi \Sigma $ and $\pi
\Sigma\to \pi \Sigma$ production mechanisms. For this purpose we have
taken the $\pi \Sigma$ and $\bar K N$ states in isospin I=0 and solved
the coupled channels Bethe Salpeter equations in terms of a potential
suggested by chiral dynamics but with free parameters. Note that the
$\pi^0 \Sigma^0$ channel has the advantage that only the isospin I=0 is
relevant and hence the analysis is simpler. We show that the fit 
determines the potential with a precision that allows one to conclude
that there are two poles, one around 1390 MeV and wide and another one
around 1420 MeV and narrow, like most chiral unitary approaches get from
the analysis of scattering data. The results of this work are most
opportune at a time when some recent fits to the scattering data are
providing different pole structures than the so far accepted by the
different theoretical groups, but which in our opinion would fail to
reproduce the results of the $\Lambda(1405)$ production data
\cite{mai,ollernew}.\footnote{In \cite{ollernew} two solutions are
proposed, one which is compatible with other results, including also
\cite{Oller:2000fj}, and another one more problematic. If one allows for
uncertainties in the normalization, a broad band of the $\pi \Sigma$
mass distributions is obtained that gives the impression of agreement
with the data, but normalized to the peak the deficiencies become more
clear.}

\section{Unitarized meson-baryon amplitude}

The main aim of the present work is to propose a way to extract, from experimental 
photoproduction data, the
information of the two $\Lambda(1405)$ poles
predicted by the chiral unitary
approach.

In the chiral unitary approach the 
 $\Lambda(1405)$ is generated dynamically from the final
state interaction of the meson-baryon pair.
The details for the construction of the
meson-baryon unitarized amplitude can be found in Refs.~\cite{Oset:1998it,
Oller:2000fj,Hyodo:2006kg,Oset:2001cn}. 
In the 
following we summarize the formalism for the sake of completeness and 
we show the way in which we allow the model to be modified to get a
better fine tuning from the fit to photoproduction data.
 
From the lowest order chiral Lagrangian for the interaction of the octet of 
Goldstone bosons with the octet of the low lying $1/2^+$ baryons
\cite{Bernard:1995dp}  the tree level transition amplitudes in $s$-wave
can be
obtained~\cite{Oset:2001cn} and give

\ba
    V_{ij}(\rts)&=&-C_{ij}\frac{1}{4f^2}(2\rts-M_i-M_j)\nn\\
    &\times&\left(\frac{M_i+E_i}{2M_i}\right)^{1/2}
    \left(\frac{M_j+E_j}{2M_j}\right)^{1/2}, 
    \label{eq:WT}
\ea
with $\rts$ the center of mass energy, $f$ the averaged meson decay
constant $f = 1.123f_\pi$~\cite{Oset:2001cn} with $f_\pi = 92.4\mev$,
$E_i$ ($M_i$) the energies (masses) of the baryons of the $i$-th 
channel and $C_{ij}$ coefficients  given, for isospin $I=0$, by 
\begin{equation} C_{ij} =\begin{pmatrix} 3 &
-\sqrt{\frac{3}{2}}  \\ -\sqrt{\frac{3}{2}}& 4 
\end{pmatrix} 
\label{eq:couplingI}. 
\end{equation}
The $i$ and $j$ subscripts represent the channels  $\bar K N$ and
$\pi\Sigma$ in isospin-basis. Note that we do not  consider  the other
possible channels in $I=0$,  $\eta\Lambda$ and $K\Xi$,  for the sake of
simplicity of the approach and because for the energies that we will
consider in this work the effect of those channels can be effectively
reabsorbed in the subtraction constants, as explained in the next
section. We are only interested in the $I=0$ channel since we will 
consider only the $\pi^0\Sigma^0$ final meson-baryon state in the
experiment of CLAS, which can only be in $I=0$ and $I=2$, but the $I=2$
is  non-resonant and negligible. In this way one has not to bother about
isospin 1 contributions which would significantly increase the complexity
of the analysis.

The implementation of 
unitarity in coupled channels of the scattering amplitude 
is one of the crucial points of the chiral
unitary approach. This can be accomplished by means of
the Inverse Amplitude
Method \cite{dobado-pelaez,Oller:1998hw} or the N/D method 
\cite{Oller:1998zr,Oller:2000fj,Hyodo:2003qa}. In this latter
work the equivalence with the Bethe-Salpeter equation
used in \cite{Oller:1997ti} was established.
 Based on the $N/D$ method, the coupled-channel scattering amplitude 
$T_{ij}$ is given by the matrix equation
\begin{equation}
   T=[1-VG]^{-1}V ,
   \label{eq:BS}
\end{equation}
where $V_{ij}$ is the interaction kernel of Eq.~(\ref{eq:WT}) and the 
function $G_{i}$, or unitary bubble, is given by the dispersion integral of 
the two-body phase space 
$\rho_i(s)=2M_{i}q_i/(8\pi W)$, in a diagonal matrix 
form,  with $M_i$ the mass of the baryon of the meson baryon loop, $q_i$ the on shell momentum of the particles of the loop and $W$ the center of mass energy. 

This $G_i$ function is  
equivalent to the meson-baryon loop function
\begin{eqnarray}
G_{i} &=& i \, \int \frac{d^4 q}{(2 \pi)^4} \, \frac{M_i}{E_i
(\vec{q}\,)} \nn\\
&\times&\frac{1}{k^0 + p^0 - q^0 - E_i (\vec{q}\,) + i \epsilon} \,
\frac{1}{q^2 - m^2_i + i \epsilon} ~.
\label{gloop}
\end{eqnarray}
The integral above
is divergent, and therefore it has to be regularized, which
can be done either with a three momentum cutoff,
or with dimensional regularization in terms of a 
subtraction constant $a_i$. The connection between
both methods was shown in Refs. \cite{Oller:1998hw,Oller:2000fj}.
In ref.~\cite{Oset:2001cn,cola}
 the values $a_{KN} = -1.84$, $a_{\pi\Sigma}=-2$
where used. In the present case, since we do not consider the 
$\eta\Lambda$ and $K\Xi$ channels these subtraction constants may differ
slightly but we will allow to vary these constants in the
fit below.

\begin{figure}[!h]
\begin{center}
\includegraphics[width=0.9\linewidth]{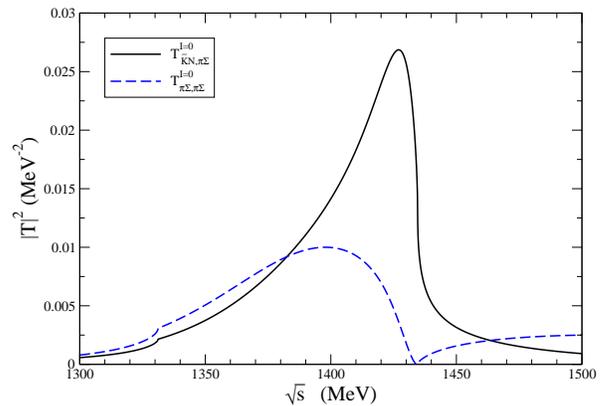}
\caption{Modulus squared of the  meson-baryon unitarized
amplitudes $T^{I=0}_{\bar K N,\pi\Sigma}$ (solid line) and 
$T^{I=0}_{\pi\Sigma,\pi\Sigma}$ (dashed line).}
\label{fig:t_MMMM}
\end{center}
\end{figure}

The amplitudes $T_{\bar K N\to\pi \Sigma}$ and $T_{\pi \Sigma\to
\pi \Sigma}$ for $I=0$ are shown in fig.~\ref{fig:t_MMMM}. They
 produce two poles in the
second Riemann sheet of the complex energy plane at the positions 
$\sqrt{s_0}=1387-67i\mev$,
and $1437-13i\mev$. Note that the poles come dynamically from the
non-linear dynamics involved in the implementation of unitarity in the
meson-baryon scattering amplitude, without
the need to include the poles as explicit degrees of freedom. This is
what is usually called {\it dynamically generated} resonance or
meson-baryon molecule.
It is worth mentioning that the unitarized amplitudes
provide the
actual meson-baryon scattering amplitudes, not only the poles of the
resonance in the complex plane. Indeed the resonant shape
of the amplitudes around the 1400~MeV region are far from a
Breit-Wigner--like shape. Therefore a fit assuming 
Breit-Wigner resonant shapes to experimental
data is not suitable for this resonance and a model like the present
one, in the line of
implementing unitarity in coupled channels, is called for in order to
reproduce or fit experimental data where these amplitudes
are relevant.

\section{Fit to photoproduction data}

In ref.~\cite{Moriya:2013eb}, data for the 
$\gamma p\to K^+ \pi^+\Sigma^-$, $\gamma p\to K^+ \pi^-\Sigma^+$  and $\gamma p\to K^+ \pi^0\Sigma^0$ reactions were taken at different photon energies.
The $\gamma p\to K^+ \pi^0\Sigma^0$ reaction filters I=0 and these are the data that we will use. 
The main observable measured for this reaction is 
the $\pi^0\Sigma^0$ invariant mass distribution  (see fig.~\ref{fig:fitVcte} below).

\begin{figure}[!h]
\begin{center}
\includegraphics[width=0.9\linewidth]{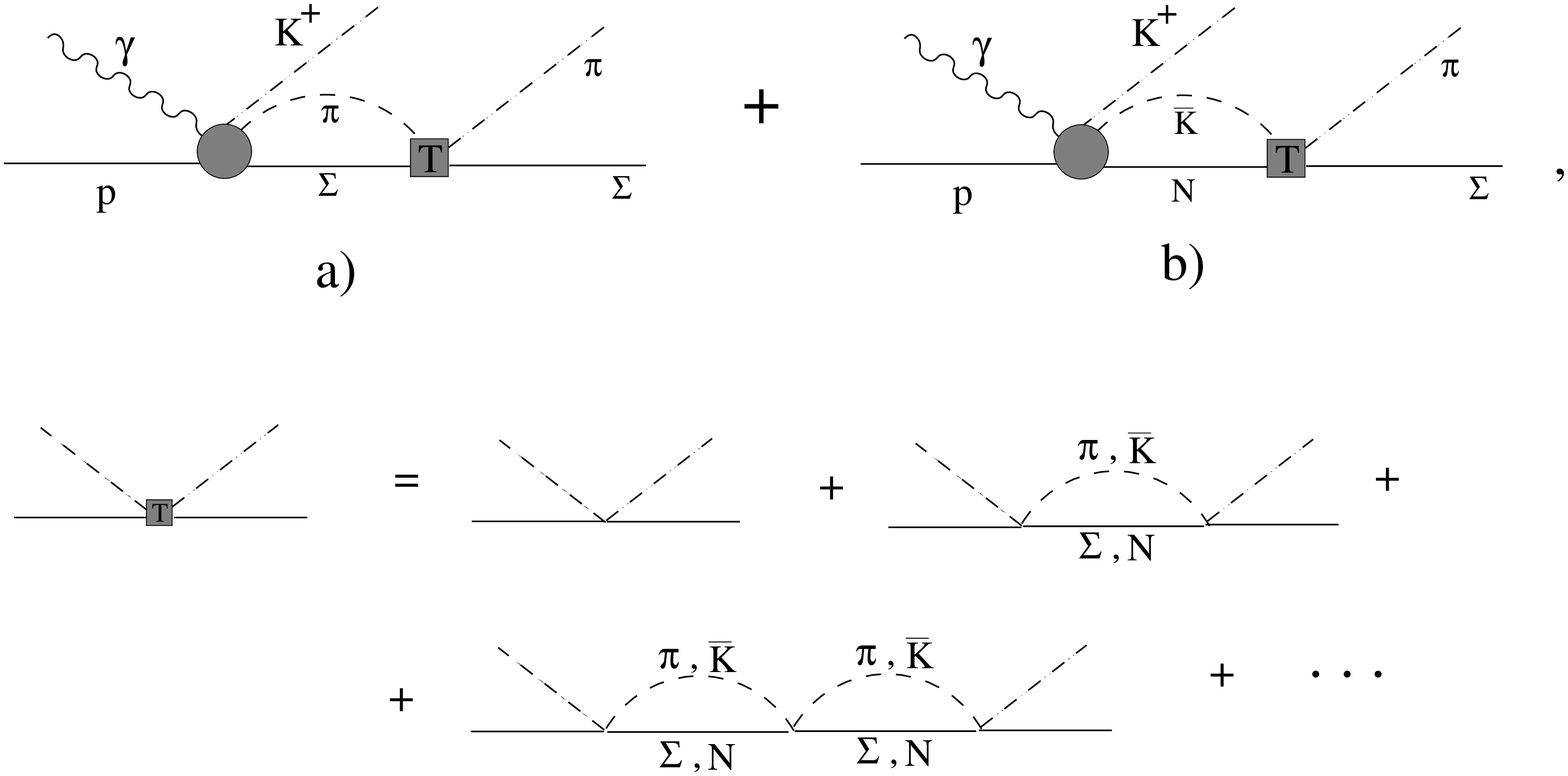}
\caption{General mechanisms for the $\Lambda(1405)$ photoproduction in  $\gamma p\to K^+
\pi^0\Sigma^0$ reaction.}
\label{fig:diagrams}
\end{center}
\end{figure}
Since the $\Lambda(1405)$ is  dynamically
generated from the final state interaction of the meson-baryon
produced, the most general mechanisms for the
photoproduction reaction are those depicted in fig.~\ref{fig:diagrams},
a) and b).
The photoproduction can proceed by the production of either a
$\pi\Sigma$, (fig.~\ref{fig:diagrams}a), 
or $\bar K N$ (fig.~\ref{fig:diagrams}b) pair, thick circle in  fig.~\ref{fig:diagrams}, 
which rescatter to produce the final $\pi\Sigma$, accounted for by the
unitarized scattering amplitude explained in previous section.
Note that a possible contact mechanism of direct $\pi\Sigma$ production
would contribute to the background and we do not consider it 
since a proper background subtraction has been done in the 
experimental analysis.

Based on fig.~\ref{fig:diagrams} 
it is immediate to realize that the
amplitude for the photoproduction process can be generally
written as
\be
t(W)=b(W) G_{\pi\Sigma} T^{I=0}_{\pi\Sigma,\pi\Sigma}
+c(W) G_{\bar K N} T^{I=0}_{\bar K N,\pi\Sigma}\ ,
\label{eq:tampli}
\ee
with $W$ the energy of the $\gamma p$ interaction.
The coefficients $b$ and $c$ may in general depend on  $W$ 
and hence we consider 9 sets of them labeled $b_j$ and $c_j$, with 
$j$ from 1 to 9, in order to
account for the 9 different energies $W$ provided
by the experimental result
of CLAS \cite{Moriya:2013eb}. On the other hand the relative weight between the 
 $G_{\pi\Sigma} T_{\pi\Sigma,\pi\Sigma}$ and 
 $G_{\bar K N} T_{\bar K N,\pi\Sigma}$ amplitudes must be complex in
 general,
 therefore we allow the $c_j$ to be complex and keep $b_j$ real since a
 global phase in the total amplitude is irrelevant.
Note that we have intentionally avoided proposing
any model for the initial
photoproduction mechanisms since we aim at suggesting a way to extract
physical poles of the $\Lambda(1405)$ resonance from experimental data
in a way as model independent as possible to ease the implementation by
experimental groups. Indeed these initial photoproduction mechanisms 
are encoded in the coefficients $b$
and $c$. Since we are fitting 9 different energies we have thus 
in total 27
parameters. This may look large but none of them affect the meson-baryon
scattering amplitude and the number is smaller than in other possible
experiments 
where mixing with isospin 1
could be allowed, like for instance
 $\gamma p \to K^+ \pi^\pm \Sigma^\mp$ in the same CLAS experiment.
 
 One has to view the fit from the perspective that the data for one
  energy will provide the three coefficients, $b$ and $c$ (complex) at this
  energy. Only the parameters of the potential affect all the data. This
  problem is similar  to the fit conducted to pionic atoms to extract
  neutron radii in \cite{neutronrad}. In that problem there were 19
  parameters for 19 neutron radii and 6 parameters for the potential.
  Again, each of these 19 parameters affected only the data on shifts
  and widths of a single pionic atom and the 6 parameters of the
  potential affected all the data. The fits worked without problems and
  the set of neutron radii obtained is considered nowadays the most
  valuable experimental source of neutron radii, together with the
  information obtained from antiprotonic atoms in \cite{pabarneutron}. 

We first  fit the $b$
and $c$ coefficients to the photoproduction $\pi^0\Sigma^0$
invariant mass distribution data
using for
the unitarized amplitudes the expression and parameters explained in the
previous section.
Note that in this first step the chiral unitary amplitudes
for the meson-baryon
interaction are kept constant (see fig.~\ref{fig:t_MMMM}).
 Only the photoproduction
vertex is allowed to vary.
The results of this fit is shown in fig.~\ref{fig:fitVcte}.
\begin{figure*}[h]
\begin{center}
\includegraphics[width=0.8\textwidth]{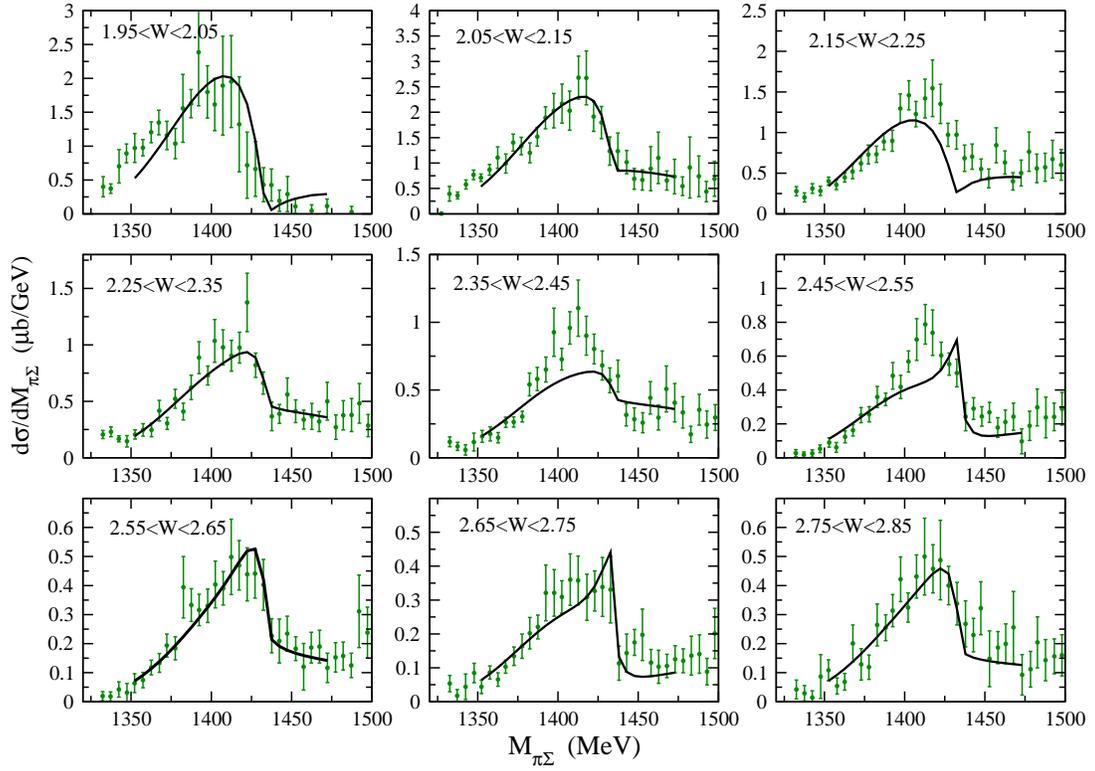}
\caption{Fit with fix unitary amplitudes, $\alpha_i=1$.}
\label{fig:fitVcte}
\end{center}
\end{figure*}
In the evaluation of the theoretical invariant mass distribution
 the three body phase space
has been averaged within the experimental $W$ bin,
 $[W-0.05,W+0.05]\gev$,
 for every W. In the fit the range $M_{\pi\Sigma}\in[1350,1475]\mev$ is
 considered.
The fit is fair for most of the energies, ($\chi^2/dof=1.76$),
 which means
that an actual full physical meson-baryon amplitudes must not be much
far from those predicted by the chiral unitary approach\footnote{
This value of the $\chi^2/dof$ is already better than the one of the
best fit in \cite{Moriya:2013eb}, $\chi^2/dof=2.15$.}.
But what we actually want in the present work is not to calculate
what the chiral unitary approach predicts for
the poles of the $\Lambda(1405)$ but to extract them from the
 experimental photoproduction 
data. Therefore we can try to get results with $\chi^2/dof\simeq
1$ by
allowing the basic chiral unitary model to vary slightly. 
In this way we
could obtain a fine tuning of the chiral unitary model and then of the
position of the $\Lambda(1405)$ poles.
In order to do this we multiply each coefficient of the 
potentials of the unitary
amplitudes,  
Eq.~(\ref{eq:couplingI}), by one parameter $\alpha_i$ and hence the new
coefficient matrix that we consider now is given by
 \begin{equation} C_{ij} =\begin{pmatrix} 3 \alpha_1 &
-\sqrt{\frac{3}{2}} \alpha_2  \\ -\sqrt{\frac{3}{2}}\alpha_2& 4 \alpha_3
\end{pmatrix} 
\label{eq:couplingI}. 
\end{equation}
Furthermore we also allow to vary the subtraction constants from
the regularization of the loop function by multiplying both of them by 
a free parameter, $\alpha_4$, $\alpha_5$:
   $a_{KN}\to \alpha_4 a_{KN}$, $a_{\pi\Sigma}\to \alpha_5
a_{\pi\Sigma}$.
Therefore, the chiral unitary amplitudes depend on 5 free parameters,
$\alpha_i$,
to be fitted and with the potential obtained we shall search for the positions of the two $\Lambda(1405)$ poles.

\begin{figure*}[h]
\begin{center}
\includegraphics[width=0.8\textwidth]{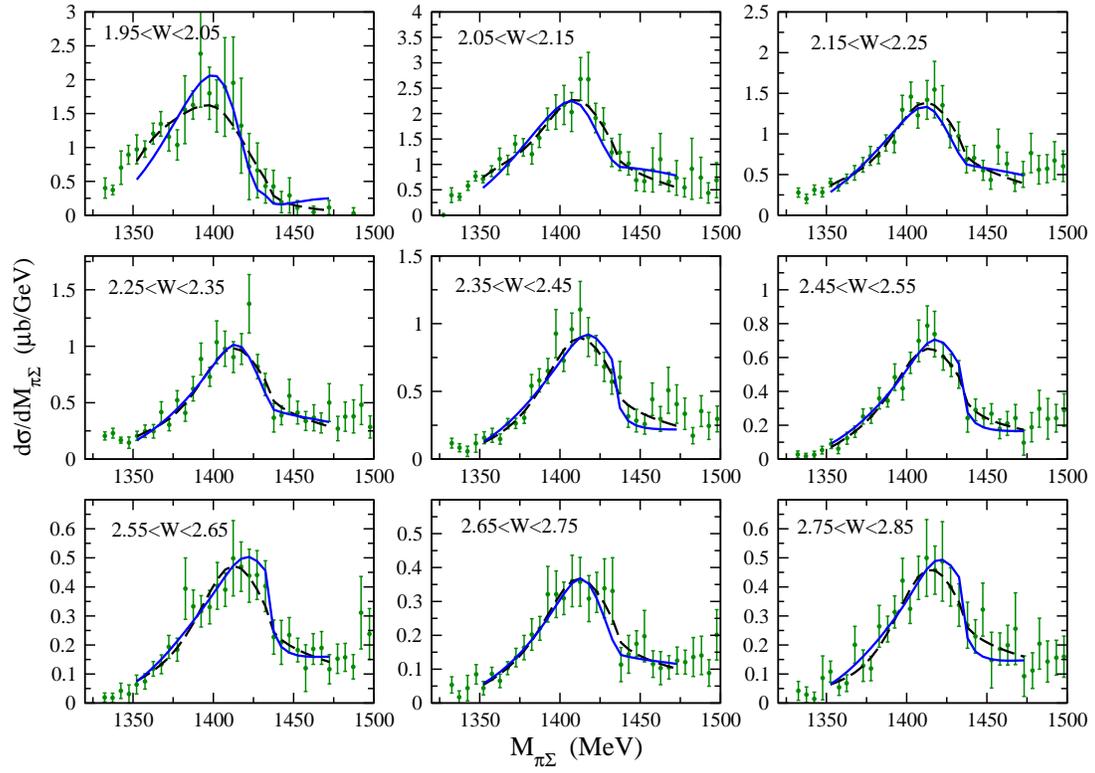}
\caption{Results from solutions 1 (solid line) and 2 (dashed line)
explained in the text.}
\label{fig:fitsols12}
\end{center}
\end{figure*}

If at this point we carry on a global fit allowing for all the parameters to be 
free from the
beginning in the
fitting algorithm, there are many local minima of the  $\chi^2$ 
function,
most of them having clearly unphysical values of
the parameters. Therefore it is very difficult to get and identify 
an absolute
minimum. Actually many 
minima have $\chi^2$ very similar but with
very different values of the parameters, which spoils the statistical
significance of the fit and the possible physical conclusions.
In order to get physically meaningful results,
we implement the following strategy:
The previous fit of fig.~\ref{fig:fitVcte}, i.e. fixing $\alpha_i=1$, is already
reasonably fair, and the potential is consistent with data of scattering \cite{Oset:1998it}, 
hence a good physical global fit
should not be very far from having values of $\alpha_i\sim 1$.
Therefore, in a first step, we start
from the fit of fig.~\ref{fig:fitVcte}, which was obtained fixing
$\alpha_i=1$, but fixing now the $b_j$ and
$c_j$ parameters and allowing only the $\alpha_i$ parameters to change. 
Next, fixing
the new $\alpha_i$ parameters obtained in the previous step,
we fit again the $b_j$ and
$c_j$ parameters and iterate the process till
 we get a $\chi^2/dof\simeq 1$ 
(which we call solution 1 in the following). After this iteration we
 get the result shown in fig.~\ref{fig:fitsols12}
  and the $\alpha_i$ parameters obtained are shown in
table~\ref{tab:results} besides the corresponding poles of
the $\Lambda(1405)$.

\begin{table}[h]
\begin{center}
\begin{tabular}{|c|c|c|c|c|c|c|c|} 
\hline 
 & $\alpha_1$ & $\alpha_2$ & $\alpha_3$& $\alpha_4$& $\alpha_5$&
 \multicolumn{2}{c|}{$\Lambda(1405)$ poles [MeV]} 
\\ \hline
 solution 1 & 1.15 & 1.17       & 1.15 & 1.03 & 0.88  & 1385-68i& 1419-22i\\ \hline
 solution 2 & 1.88 & 1.89       & 1.57 & 0.93 & 0.87  & 1347-28i& 1409-33i\\ \hline
 \end{tabular}
\end{center}
\caption{Parameters of the unitarized amplitudes and pole positions of
the $\Lambda(1405)$ for
both solutions discussed in the text}
\label{tab:results}
\end{table}

 We can see that the parameters obtained 
are not very different from 
1 for solution 1. This means that allowing for a small variation in the 
parameters of the chiral unitary approach the photoproduction data can
be properly reproduced. In other words, a
 small freedom in changing
the $\alpha_i$ coefficients allows for extracting
the pole positions for the two $\Lambda(1405)$ from experimental
photoproduction data. The results obtained for the poles 
are  $1385-68i\mev$ and $1419-22i\mev$.
The solution 1 shown above 
does not actually correspond to the minimum $\chi^2/dof$ but
to a $\chi^2/dof\simeq 1$. The absolute minimum that we find
after iterating the process described above many times has
$\chi^2/dof=0.60$ (solution 2) but since it is smaller than 1 it is not more
statistically significant that the one with $\chi^2/dof\simeq 1$
(solution 1).
The fit for photoproduction for solution 2 is represented by the
dashed-line in fig.~\ref{fig:fitsols12}
and the coefficients and corresponding
 poles in table~\ref{tab:results}.
The $\alpha_i$ coefficients for solution 2
differ more from the chiral unitary approach predictions,
($\alpha_i=1$), than those from solution 1.
The pole positions obtained with
these parameters are not far
from those of solution 1. Anyway we
could consider the difference between solutions 1 and 2 
as a conservative estimate of the 
uncertainties of the procedure. Yet, a visual inspection to  fig.~\ref{fig:fitsols12} induces us to accept the solution 1 as better than solution 2 because it respects much better the Flatt\'e behaviour (with a fast fall of the cross section in the upper side of the mass distribution) exhibited by the experimental data.

\begin{figure}[h]
\begin{center}
\includegraphics[width=0.95\linewidth]{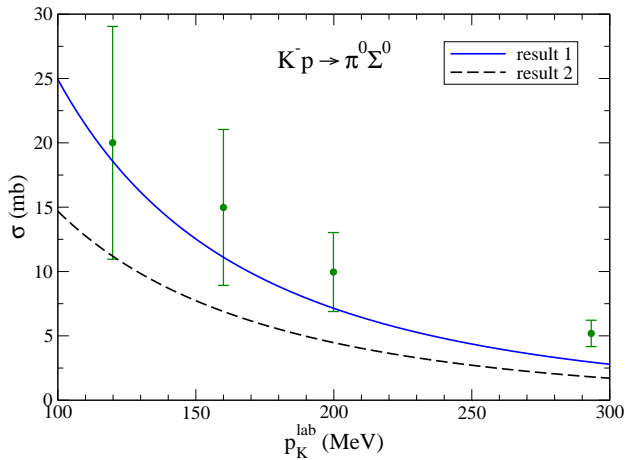}
\caption{Predicted $K^-
p\to\pi^0\Sigma^0$ cross section for solutions 1 and 2. Experimental
data from refs.~\cite{Humphrey:1962zz,kim}.}
\label{fig:crosskp}
\end{center}
\end{figure}

In order to make further checks that the fits obtained are physically
acceptable and to decide between the different solutions obtained above,
we  calculate now the cross section for $K^-
p\to\pi^0\Sigma^0$ interaction which is the only one that does not mix
with $I=1$. The amplitude for this reaction is purely $I=0$, which is the isospin involved in
the fit to photoproduction data, (since the $I=2$ is negligible).
The result is shown in fig.\ref{fig:crosskp} in comparison to experimental data
from refs.~\cite{Humphrey:1962zz,kim}.
It can be seen that the best result corresponds to solution 1. Note,
however, that as the energy increases other channels like
 $\eta\Lambda$ and $K\Xi$ are needed to be explicitly included as well
as higher order contributions in the chiral potentials, but for low energies, and particularly to determine the position of the poles of the $\Lambda(1405)$, our analysis, with the channels chosen and the freedom of the potential is sufficient. 

Since we have been concerned about the poles of the $\Lambda(1405)$, our analysis using only the I=0 data is appropriate. In the future one can think of
using also the $\gamma p \to K^+ \pi^+ \Sigma^-$ and 
$\gamma p \to K^+ \pi^- \Sigma^+$ data to try to induce the I=1 potential. In this case one can also use the large set of $K^- p$ scattering data to
 constrain further
the potential. A global fit to all the data and using potentials beyond the lowest order would certainly be most welcome and one could hopefully determine whether there is or not an I=1 state around 1430 MeV, which has been hinted in 
\cite{Oller:2000fj} and \cite{cola}
and also obtained by the fit of \cite{Moriya:2013eb}.

One should note that the global fit obtained in \cite{Moriya:2013eb} is
admittedly rather imperfect and, as quoted there, the authors are unable
to get a reduced $\chi^2$ smaller than 2.15. Instead we get fits of high
quality with the reduced $\chi^2$ of the order of 1. Furthermore, an
inspection of the results of our fit in Fig.~\ref{fig:fitsols12} and 
those of Fig.~21 of \cite{Moriya:2013eb} for $\gamma p \to K^+ \pi^0
\Sigma^0 $ clearly shows that the fit to the data is much better in our
analysis.

It is interesting to see why our fit to the data of \cite{Moriya:2013eb}
is better than the one obtained in this latter work. There is an essential difference
between our analysis and the one of \cite{Moriya:2013eb}. 
In \cite{Moriya:2013eb} the $\gamma p \to K^+ \pi
\Sigma$ amplitudes are parametrized as (Eq.(19) of \cite{Moriya:2013eb})
\be
t_I(m)=C_I(W)e^{i\Delta\phi_I}B_I(m),
\ee
where $C_I(W)$ is a weight factor, $\Delta\phi_I$ a phase and $B_I(m)$ a
Breit-Wigner function. As one can see, the weight is allowed to depend
on the photon energy, $W$, but not its phase. But even more restrictive is
the fact that the shape of the resonance, $B_I(m)$, is chosen independent of the
photon energy. This neglects the possibility that one has two poles of
the $\Lambda(1405)$ resonance and that the amplitudes $\gamma p \to K^+ \pi
\Sigma$ are superpositions of the amplitudes corresponding to these
poles with relative weights that depend on the photon energy. Since this
is what happens in the theories that predict two poles, it is then
important that an analysis of the data takes this into account and this
is done in our analysis. In our analysis the amplitude is given by
Eq.~(\ref{eq:tampli}) as a superposition of the 
$T^{I=0}_{\pi\Sigma,\pi\Sigma}$ and
$T^{I=0}_{\bar K N,\pi\Sigma}$ amplitudes, which have a very different
shape as seen in Fig.~\ref{fig:t_MMMM}. With $b(W)$ and $c(W)$ depending
on the photon energy, we allow the freedom to change the shape of the
resonance as the photon energy changes. 
\\

\section{Conclusions} 

We have studied the  $\gamma p \to K^+ \pi^0 \Sigma^0 $ reaction at
different energies from a semiempirical point of view in order to
illustrate the possibility to obtain the position of the two
$\Lambda(1405)$ poles from experimental production data. We have taken an
amplitude for this reaction which consists of a linear combination of
the $\pi \Sigma \to \pi \Sigma$ and $\bar K N \to \pi \Sigma$ scattering
amplitudes in I=0. The parameters of this combination are free and
different for each energy. The $\pi \Sigma \to \pi \Sigma$ and $\bar K N
\to \pi \Sigma$ amplitudes are constructed using the Bethe Salpeter
equations in the coupled channels of $\bar K N$ and $\pi \Sigma$ in
isospin I=0. For this we used a $2 \times 2$ potential matrix inspired
by the chiral unitary approach but slightly modified 
with free coefficients. These
coefficients and those of the linear combinations were fitted to the
data and good solutions were obtained. It was interesting to see that
these solutions gave fair results for the cross section of the $K^- p
\to \pi^0 \Sigma^0 $ reaction and provided two poles very close to those
provided by the chiral unitary approaches. By choosing the set of
parameters that also provides best results for the $K^- p \to \pi^0
\Sigma^0 $ reaction we could decide the best results 
and we found for the two poles of the $\Lambda(1405)$
$1385-68i\mev$
and $1419-22i\mev$ in the complex energy plane. The exercise
conducted in the present work is orthogonal and complementary to the usual one
performed so far where the scattering data are used to fit the
parameters of the chiral theory. We find that the photoproduction data
at several energies have enough information to provide the
$\Lambda(1405)$ poles, without the need to develop a detailed model for
the reaction. After this work, two lines of progress look most
advisable: Elaborating a detailed model to describe the data
theoretically and performing simultaneous fits to the scattering data
and  $\Lambda(1405)$ production data. The present work has shown
clearly that the information contained in the $\Lambda(1405)$ production
data is extremely valuable to learn about the position of the
$\Lambda(1405)$ poles and the nature of these states.

 \section*{Acknowledgments}
  This work is partly supported by the Spanish Ministerio de Economia y Competitividad and European FEDER funds under the contract number
FIS2011-28853-C02-01, and the Generalitat Valenciana in the program Prometeo, 2009/090. We acknowledge the support of the European Community-Research Infrastructure Integrating Activity
Study of Strongly Interacting Matter (acronym HadronPhysics3, Grant Agreement
n. 283286) under the Seventh Framework Programme of EU.

\end{document}